# Bitcoin Under Stress: Measuring Infrastructure Resilience 2014–2025

Wenbin Wu
*Cambridge Centre for Alternative Finance*
Cambridge, UK
w.wu@jbs.cam.ac.uk

Alexander Neumueller
*Cambridge Centre for Alternative Finance*
Cambridge, UK
a.neumueller@jbs.cam.ac.uk

***Abstract*—Bitcoin's design promises resilience through decentralization, yet the physical infrastructure supporting the network creates hidden dependencies. We present the first longitudinal study of Bitcoin's resilience to submarine cable failures, using 11 years of P2P network data (2014–2025) and 68 verified cable fault events. Applying a Buldyrev-style cascade model at country level, we find that Bitcoin's clearnet (non-Tor) critical failure threshold $p_c \approx 0.72$–$0.92$ for random failures, meaning the vast majority of inter-country cables must fail before significant node disconnection. Targeted attacks are an order of magnitude more effective ($p_c = 0.05$–$0.20$). To address the majority of nodes now using Tor with unobservable locations, we develop a 4-layer multiplex model incorporating Tor relay infrastructure. Because relay bandwidth concentrates in well-connected European countries, Tor adoption *increases* resilience under current relay geography ($\Delta p_c \approx +0.02$–$+0.10$) rather than introducing hidden fragility. Empirical validation confirms weak physical-layer coupling: 87% of historical cable faults caused less than 5% node impact. We contribute: (1) a multiplex percolation framework for overlay-underlay coupling, including a 4-layer Tor relay model; (2) the first empirical measurement of Bitcoin's physical-layer resilience over a decade; and (3) evidence that Tor adoption amplifies resilience, with distributional bounds quantifying uncertainty under partial observability.**

## I. Introduction

[1]

Bitcoin represents a natural experiment in distributed systems resilience. Its peer-to-peer network has operated continuously since 2009, surviving regulatory actions, infrastructure failures, and targeted attacks [1], [2]. Yet Bitcoin's resilience claims rest primarily on protocol-level guarantees (e.g., difficulty adjustment, 51% attack resistance) rather than systematic analysis of physical infrastructure dependencies.

We address a fundamental question: *How resilient is Bitcoin to physical infrastructure disruption?* Unlike prior work focused on protocol attacks [3] or mining concentration [4], we analyze the coupling between Bitcoin's logical overlay and physical underlay, from submarine cables through BGP routing to P2P node connectivity.

Our work extends the theory of interdependent networks [5] to systems with *structural redundancy* and *partial observability*. Buldyrev et al. showed that tightly coupled networks exhibit catastrophic cascading failures; we find that Bitcoin's multi-homed architecture produces the opposite: physical layer failures rarely cascade to network fragmentation.

We make three contributions. First, we develop a *multiplex percolation framework* for overlay-underlay coupling, extending Buldyrev-style cascades [5] with the tensorial approach of Kivelä et al. [6], including a 4-layer model that incorporates Tor relay infrastructure as a distinct network layer. Second, we provide the *first empirical measurement* of Bitcoin's physical-layer resilience over a decade, using 658 submarine cables and verified fault events. Third, we show that *Tor adoption amplifies resilience* due to relay concentration in well-connected countries, with distributional bounds quantifying uncertainty under partial observability.

Bitcoin is highly resilient to random cable failures but an order of magnitude more vulnerable to targeted attacks. Clearnet resilience declined during 2018–2021 as the network grew rapidly and mining concentrated geographically, then partially recovered. Contrary to the concern that Tor's growing adoption introduces unquantifiable risk, our 4-layer model shows that Tor *strengthens* resilience: relay infrastructure concentrates in well-connected European countries, creating a compound barrier to disruption that consistently exceeds the clearnet-only baseline.

## II. Background and Related Work

### A. Bitcoin P2P Network Architecture

Bitcoin's P2P network consists of *nodes* that validate transactions and maintain consensus. Nodes discover peers through DNS seeds and addr propagation, maintaining connections to typically 8–125 other nodes [7]. Network measurement studies have characterized topology [8], [9], [10], finding that the network maintains robustness despite heavy-tailed degree distributions.

The network operates as an *overlay* on physical internet infrastructure. Each node has an IP address (clearnet) or Onion address (Tor), connecting through ISPs, IXPs, and submarine cables. This creates a dependency chain from physical to logical layers. Recent work has begun mapping these dependencies in the general Internet: Ramanathan and Abdu Jyothi [11] map IP links to submarine cables with 91% coverage, and Xaminer [12] traces how physical-layer failures cascade through IP routing. However, no prior study applies

[1]We thank Y.-Y. Ahn, A. Arenas, U. Bindseil, R. de Oliveira Costa, C. Diaz, G. Dong, B. Gross, M. Grundmann, H. Halaburda, S. Havlin, R. Jansen, J. Lopp, O. Maennel, F. Radicchi, A. Ramanathan, L. Shekhtman, N. Starosielski, P. Syverson, and F. Tschorsch for helpful comments on an earlier draft.

such cross-layer mapping to Bitcoin or validates against cable fault events.

Prior longitudinal studies have characterized Bitcoin's P2P overlay topology [13] and AS-level distribution [2], [14], but none maps to physical cable infrastructure or validates against cable fault events. Prior percolation applications to cryptocurrency networks include Bartolucci et al.'s Lightning Network model [15], which uses single-layer bond percolation for economic emergence, and Shekhtman et al. [16], who apply link percolation to the transaction graph for deanonymization analysis. Neither addresses *interdependent* cascading failures across physical-logical layers, specifically the Buldyrev-style multi-layer cascade we model here.

### B. Infrastructure Attacks on Bitcoin

Heilman et al. [1] demonstrated *eclipse attacks* where an adversary monopolizes a victim's connections. Apostolaki et al. [2] showed that BGP routing attacks could partition the network, and proposed SABRE [17] as a relay network defense. Tran et al. [3] introduced stealthier Erebus attacks exploiting AS-level topology.

These attacks assume adversarial control of routing infrastructure. Beyond adversarial scenarios, Abdu Jyothi [18] analyzed submarine cable vulnerability to solar superstorms, showing that intercontinental cables, particularly high-latitude North America–Europe routes, are the primary failure mode for large-scale internet disruption. Our work shares the focus on physical cable infrastructure but differs in scope: we study *ongoing* cable faults rather than extreme solar events, and analyze their propagation specifically through Bitcoin's overlay network rather than the general internet.

### C. Interdependent Network Theory

Buldyrev et al. [5] established that interdependent networks with *strong* coupling exhibit catastrophic cascading failures. When networks A and B have one-to-one dependency links, removing nodes from A triggers failures in B, which cascade back to A. The critical failure threshold $p_c$ depends on the degree distributions and coupling strength.

Gao et al. [19] generalized this to arbitrary *networks of networks* with $n$ interdependent layers, providing the analytical foundation for multi-layer cascade analysis. Subsequent work further extended this framework: Gao et al. [20] identified universal resilience patterns allowing dimension reduction. Osat et al. [21] analyzed optimal attack strategies on multiplex networks. Kivelä et al. [6] provided a tensorial framework for arbitrary multiplex topologies.

We adapt this framework as an *operational simulation* rather than a statistical physics model. Our physical layer models 225 countries connected by 354 submarine cable edges and 325 land border edges, capturing both intercontinental chokepoints and continental connectivity. Inter-layer dependencies arise from geographic co-location rather than arbitrary dependency links, and Bitcoin's multi-homed ASN topology creates natural redundancy that weakens cascade propagation.

## III. Data and Methodology

### A. Data Sources

We assembled a multi-layer dataset spanning 11 years (Table I).

TABLE I: Dataset summary. [†]658 cables from TeleGeography [22]: 570 in-service, 81 planned/under-construction, 7 decommissioned with geometry retained.

| Data Source | Records | Coverage |
|---|---|---|
| Bitcoin Nodes (Bitnodes) | 8M+ obs. | 2014–2025 |
| CAIDA AS Relationships | 5.7M edges | 2014–2025 |
| Submarine Cables[†] | 658 cables | Current |
| Cable Fault Events | 385 events | 2013–2025 |
| IODA Outage Detection | Curated | 2018–2025 |
| Tor Relay Metadata (Onionoo) | 9,793 relays | Current |
| Mining Geography (CBECI) | Monthly | 2019–2022 |
| Mining Geography (Hashrate Index) | Quarterly | Q1 2025–Q1 2026 |
| Bitcoin Price | Daily | 2014–2025 |

Bitnodes crawls *reachable listening nodes* that accept inbound connections, which represent approximately 40% of the estimated full network [23]. Independent monitoring by Grundmann and Hartenstein [24], with daily snapshots since 2015 including ASN mapping, corroborates Bitnodes' geographic and AS-level distributions. Our analysis therefore measures resilience of this observable subset. Unreachable nodes behind NAT or firewall depend on the same physical infrastructure, so their resilience profile does not differ qualitatively, though the absolute node counts we report are lower bounds. Moreover, unreachable nodes are predominantly behind residential NAT and carrier-grade NAT [25], and are therefore likely more geographically and AS-diverse than the cloud-concentrated reachable population [14]. Including them would increase geographic redundancy, making our $p_c$ estimates conservative lower bounds.

We verify events following event study best practices [26], using a tiered approach. Of 385 SubTel Forum fault reports, 68 unique incidents match IODA or BGP signatures. Cable events and bitnodes snapshots are matched within 7-day windows: this accounts for SubTel fault reporting delays of typically 1–3 days, Bitnodes snapshot gaps of up to 48 hours in early years, and captures the acute impact phase before cable repair, typically 5–14 days [27]. Using 3-day or 14-day windows changes the matched count by $\pm 4$ events without materially affecting impact distributions. Our primary analysis uses these verified events.

### B. Multiplex Network Model

Following Kivelä et al. [6], we model Bitcoin's infrastructure as a 3-layer multiplex $\mathcal{M} = (V_M, E_M, L)$:

$$\boldsymbol{A} = \begin{pmatrix} A^{[1]} & C^{[1,2]} & \cdots \\ C^{[2,1]} & A^{[2]} & \cdots \\ \vdots & \vdots & \ddots \end{pmatrix} \tag{1}$$

where $A^{[\alpha]}$ is the intra-layer adjacency matrix and $C^{[\alpha,\beta]}$ encodes inter-layer coupling.

Our three layers are: $L_1$ (Physical), consisting of 225 countries connected by submarine cables and land borders; $L_2$ (Routing), ASNs connected by BGP peering from CAIDA data; and $L_3$ (Network), Bitcoin nodes hosted on ASNs.

We model physical connectivity at country level: 225 countries connected by 354 inter-country submarine cable edges, derived from reverse-geocoding TeleGeography cable endpoints, and 325 land border edges from geographic adjacency data [28]. Land border edges represent terrestrial fiber connectivity that is not vulnerable to submarine cable cuts and are *never* removed during simulation. Only submarine cable edges are subject to random removal, matching the failure model where undersea cables are the primary physical failure mode. This country-level granularity captures both intercontinental chokepoints and intra-continental connectivity, avoiding the loss of resolution inherent in regional aggregation.

Unlike Buldyrev's binary one-to-one dependencies, our cascade model uses *geographic* coupling for inter-layer links: nodes in a disconnected country fail regardless of whether their ASN has presence elsewhere. This reflects network reachability: a node's connectivity depends on its physical location, not on its provider's global footprint.

We extend this 3-layer model with a 4-layer variant that adds a *Tor relay infrastructure* layer between routing and the Bitcoin P2P layer. The extended multiplex $\mathcal{M}_4$ comprises: $L_1$ (Physical), $L_2$ (Routing), $L_3$ (Tor Relay), and $L_4$ (Bitcoin P2P). Tor relay data from Onionoo [29] provides 9,793 running relays mapped to ASNs and countries by consensus weight (CW), a bandwidth-weighted measure of relay importance for circuit formation. This layer captures a dependency that the 3-layer model ignores: Bitcoin Tor nodes require guard, middle, and exit relays to form circuits, and these relays run on physical servers connected to the same submarine cables.

### C. Percolation Simulation

We implement a cascade simulation inspired by Buldyrev et al. [5], adapted for Bitcoin's architecture. Our model departs from the original framework in two key ways: (1) coupling is *many-to-one*, with many nodes per country, rather than one-to-one [30]; and (2) the physical layer represents geographic connectivity across 225 countries rather than matching the size of the logical layer. These departures are motivated by Bitcoin's actual architecture, since nodes in the same country share physical infrastructure, and the country-level physical graph captures realistic connectivity patterns. They are also consistent with established theoretical results: reducing coupling strength from full one-to-one dependency shifts the percolation transition from first-order (discontinuous) to second-order (continuous) [31], and spatially embedded interdependent networks exhibit continuous transitions rather than the catastrophic collapse predicted for random graphs [32]. The cascade proceeds in four stages: (1) remove fraction $p$ of submarine cables (random or targeted); (2) find connected components among countries to identify fragmentation (land borders remain); (3) nodes in non-main-component countries become unreachable; (4) check ASN routing graph connectivity among remaining nodes.

The critical failure threshold $p_c$ is the minimum $p$ such that $>10\%$ of nodes become disconnected. We use 10% rather than the traditional giant component collapse criterion [5] because it is more operationally meaningful for Bitcoin: at 10% node loss, block propagation latency increases measurably [10], and mining pools in disconnected countries risk producing orphan blocks. Varying this threshold from 5% to 20% shifts $p_c$ by $\pm 0.06$–0.12 across years, confirming that qualitative findings are robust to this choice. We run 1,000 Monte Carlo trials per removal fraction with 2% step sizes, yielding smooth percolation curves with standard errors $<0.006$.

Our model makes two opposing assumptions: country-level failure propagates to all nodes in that country, which is aggressive, but ASN routing provides alternative paths for connected countries, which is optimistic. These biases partially cancel; sensitivity analysis shows $p_c$ varies $\pm 0.05$ across reasonable parameter choices. Relaxing the cascade model from "all nodes fail" to "50% of nodes in disconnected countries fail" increases $p_c$ by 0.04–0.06.

Following Osat et al. [21], we compare four attack strategies: random, degree-targeted, betweenness-targeted, and ASN-targeted.

### D. Tor and Partial Observability

A critical methodological issue: as of 2025, 64% of Bitcoin nodes use Tor, making their physical location *unobservable*. We address this through two complementary approaches.

**Distributional bounds analysis.** We compute $p_c$ under five assumptions about Tor node geography, each applied to the full Buldyrev cascade: (1) *clearnet only*, baseline ignoring Tor nodes; (2) *proportional*, Tor nodes follow clearnet country distribution; (3) *uniform*, Tor nodes spread evenly across all countries with Bitcoin nodes; (4) *clustered*, Tor mirrors known Tor relay distribution by country (DE 30%, US 15%, FR 10%, NL 8%, rest distributed) [33]; (5) *worst case*, all Tor concentrated in the least-connected country by submarine cable degree. This bounds $p_c$ without requiring knowledge of true Tor node locations.

**4-layer mechanistic model.** Rather than assuming a Tor distribution, we model Tor relay infrastructure as a separate network layer. Tor relays are physical servers with known ASNs and countries [29]. When submarine cables fail, ASNs in disconnected countries lose connectivity, and Tor relays on those ASNs go offline. If the surviving relay consensus weight drops below 50% of the network total, Tor circuit formation is severely degraded and Tor-connected Bitcoin nodes lose connectivity. This captures a dependency that distributional assumptions miss: Tor nodes are not randomly placed, but depend on relay infrastructure that is itself geographically concentrated.

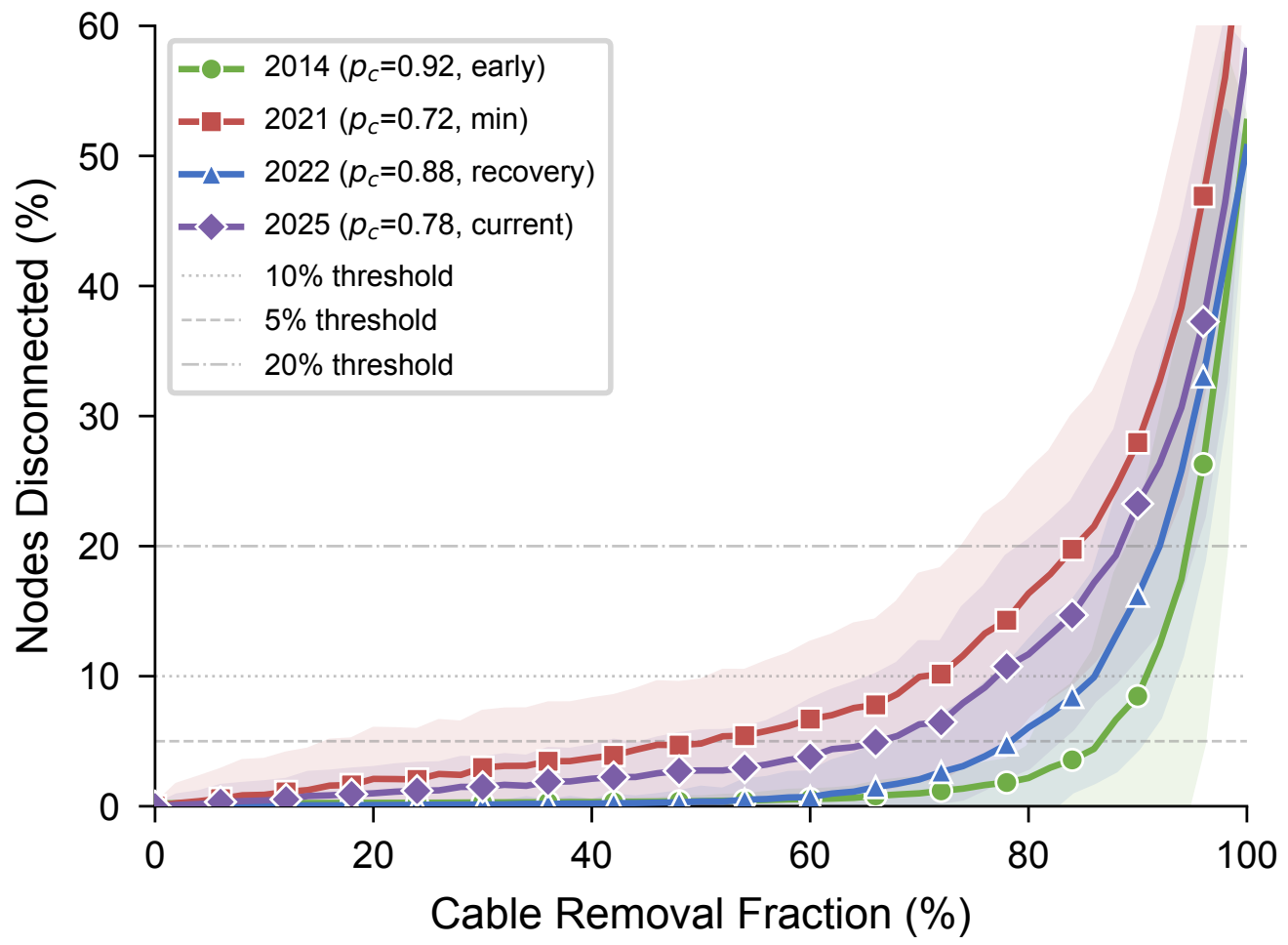

Fig. 1: Percolation curves for representative years. Shaded bands: ±1 s.d. across 1,000 trials. Horizontal lines: 5%, 10%, 20% disconnection thresholds.

## IV. Results

### A. Percolation Threshold Evolution

Table II shows critical failure thresholds from 2014–2025:

TABLE II: Temporal evolution of clearnet critical failure threshold $p_c$ (1,000 Monte Carlo trials). Nodes = clearnet nodes with valid ASN and country mapping; raw totals in Table IV. ‡Hashrate Index [34] data (195 countries); all other mining entries from CBECI [35] (10 countries).

| Year | Nodes | Countries | ASNs | Edges | $p_c$ | Mining |
|---|---|---|---|---|---|---|
| 2014 | 6,748 | 89 | 1,246 | 177K | 0.92 | – |
| 2015 | 6,260 | 82 | 1,159 | 461K | 0.92 | – |
| 2016 | 5,636 | 83 | 1,026 | 483K | 0.90 | – |
| 2017 | 5,253 | 83 | 914 | 470K | 0.92 | – |
| **2018** | **11,442** | **99** | **1,479** | **400K** | **0.82** | **–** |
| 2019 | 10,051 | 100 | 1,247 | 519K | 0.80 | E.Asia 74% |
| 2020 | 9,149 | 92 | 1,172 | 447K | 0.80 | E.Asia 65% |
| **2021** | **8,508** | **97** | **1,130** | **447K** | **0.72** | **N.Am 30%** |
| 2022 | 7,108 | 88 | 875 | 470K | 0.88 | N.Am 44% |
| 2023 | 6,589 | 91 | 962 | 506K | 0.76 | – |
| 2024 | 6,234 | 95 | 946 | 599K | 0.80 | – |
| 2025 | 7,511 | 90 | 1,048 | 689K | 0.78 | N.Am 36%‡ |

The percolation curves (Fig. 1) show distinct S-shapes across eras. The early period (2014–2017) shows high resilience at $p_c \approx 0.90$–$0.92$. A decline in 2018 marks the transition to a growth/concentration period (2018–2021), where $p_c \approx 0.72$–$0.82$, reaching the series minimum of $p_c = 0.72$ in 2021 during peak mining concentration. The recent period (2022–2025) shows partial recovery ($p_c \approx 0.76$–$0.88$): a sharp rebound in 2022 ($p_c = 0.88$) following the China mining ban is followed by renewed decline.

CBECI [35] data shows 74% of hashrate concentrated in East Asia during 2019. By 2025, Hashrate Index [34] data across 195 countries shows mining has substantially diversified: North America holds 36%, Russia 16%, China 14%, with emerging mining nations (Paraguay 4%, Ethiopia 3%, UAE 4%) distributing hashrate more broadly. Despite this diversification, $p_c$ shows no clear corresponding trend ($p_c = 0.80$ in 2019 vs. 0.78 in 2025), suggesting that mining geography and node infrastructure resilience may be largely independent—$p_c$ is driven by physical cable topology and node placement rather than hashrate distribution. However, the available mining data covers only 5 of 12 study years, limiting formal inference; the co-occurrence of peak mining concentration and minimum node resilience in 2021 suggests that geographic clustering across infrastructure layers may matter during extreme concentration events. A selection effect compounds this: as Tor adoption grew from near-zero to 64%, geographically diverse operators shifted to Tor, leaving the clearnet sample artificially concentrated. Part of the observed $p_c$ variation therefore reflects changing sample composition rather than true resilience changes in the full network.

The 2017–2018 decline coincides with a near-doubling of node count from 5,253 to 11,442, and the geographic distribution of these new nodes, concentrated in fewer countries, explains the $p_c$ drop from 0.92 to 0.82. Both rapid network growth and peak mining concentration contribute to this decline, though their relative weights are not separable in our data. Relay node geographic concentration reduced clearnet resilience by 22% from peak to trough during this period (Fig. 2).

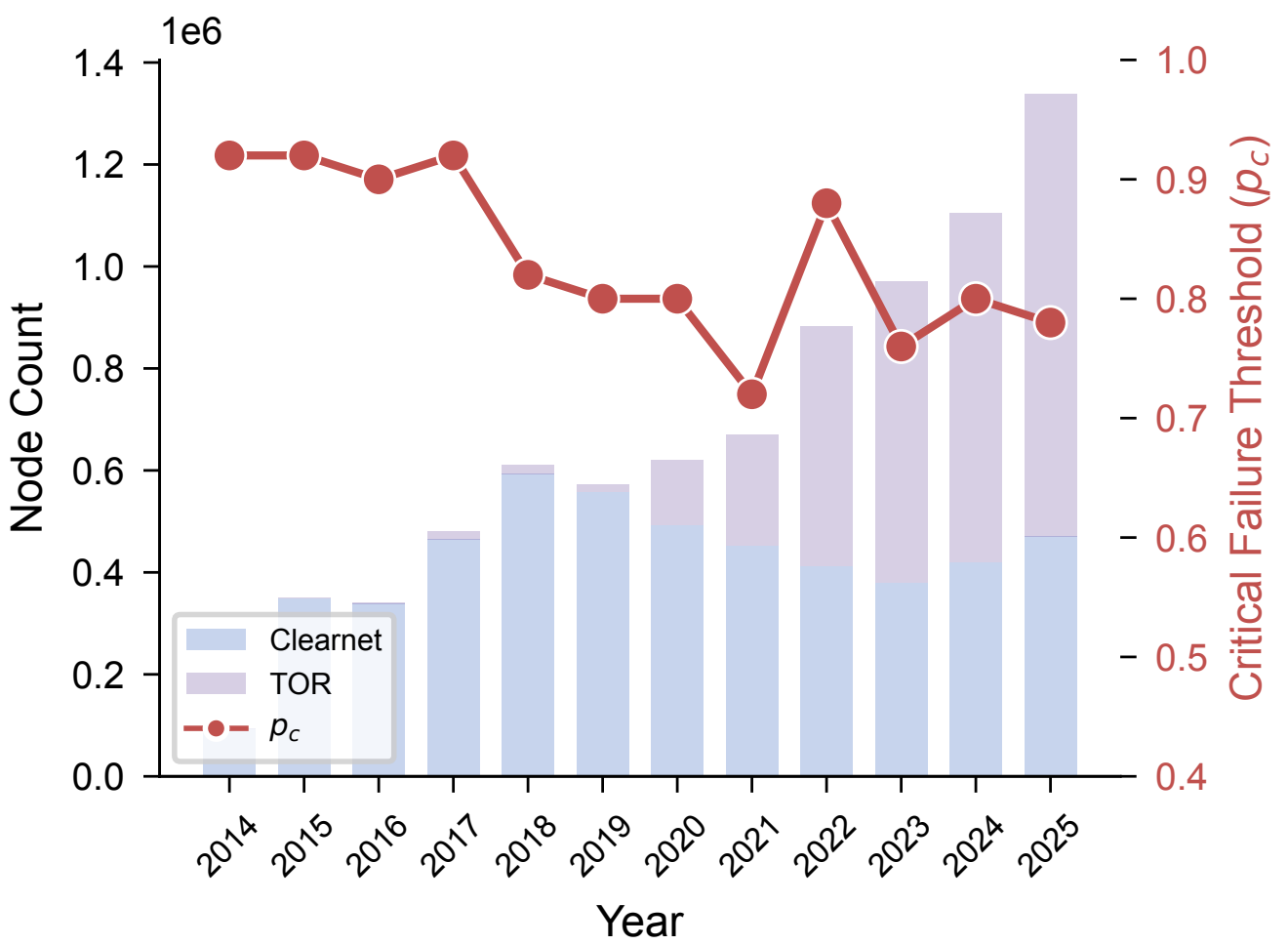

Fig. 2: Temporal evolution of network resilience (2014–2025).

### B. Random vs. Targeted Attacks

Targeted attacks are dramatically more effective than random failures (Table III, Fig. 3).

TABLE III: Attack strategy comparison. Cable attacks ($L_1$): fraction of cables removed; ASN attacks ($L_2$): fraction of routing capacity removed.

| Attack Strategy | Layer | $p_c$ | Metric |
|---|---|---|---|
| Random cable removal | $L_1$ | 0.72–0.92 | Fraction of cables |
| Targeted: high-degree cables | $L_1$ | 0.45 | Fraction of cables |
| Targeted: high-betweenness cables | $L_1$ | **0.20** | Fraction of cables |
| Targeted: top-node ASNs | $L_2$ | **0.05** | Fraction of ASN capacity |

The most critical cables by betweenness centrality are the 11 Europe–North America cables with centrality 0.141, followed by the 2 Africa–South America cables at 0.132 and the 2 Southeast Asia–South Asia cables at 0.120.

The ASN-targeted result ($p_c = 0.05$) operates on a fundamentally different attack surface than cable removal: it measures the fraction of *routing capacity* rather than cables that must be removed, targeting the top 5 ASNs by node count: Hetzner, OVH, Comcast, Amazon, and Google Cloud. This threat model maps to hosting provider shutdowns or coordinated regulatory action across jurisdictions, not physical cable cuts. However, Tor's majority share provides a floor: even complete clearnet removal leaves the majority of nodes operational.

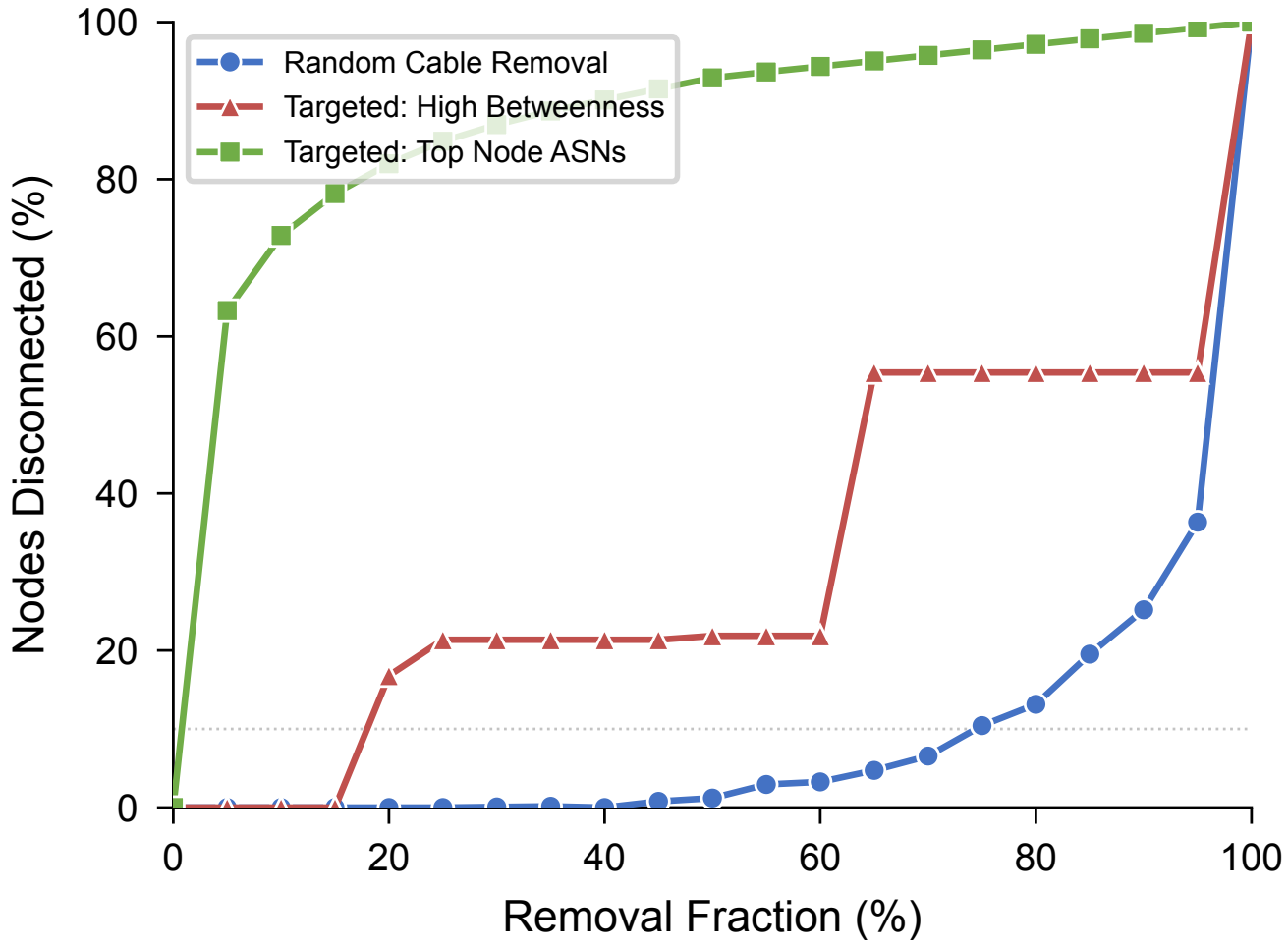

Fig. 3: Random vs. targeted attack percolation curves.

### C. Tor Adoption as Adaptive Resilience

Table IV documents the Tor adoption surge:

TABLE IV: Network composition evolution. Tor adoption surged 2021–2022, coinciding with the China mining ban.

| Year | Total | Clear | Tor | Clear% |
|---|---|---|---|---|
| 2015 | 6,289 | 6,263 | 0 | 100% |
| 2017 | 5,459 | 5,260 | 184 | 96% |
| 2019 | 10,323 | 10,103 | 216 | 98% |
| **2021** | **11,029** | **8,510** | **2,478** | **77%** |
| 2022 | 14,727 | 7,110 | 7,617 | 48% |
| 2024 | 17,197 | 6,235 | 10,962 | 36% |
| 2025 | 20,766 | 7,512 | 13,254 | 36% |

The Tor surge coincides with major censorship events: Iran shutdown (2019), Myanmar coup (2021), and China mining ban (2021). This pattern suggests *adaptive self-organization*: the Bitcoin community shifted toward censorship-resistant infrastructure without central coordination. Tor adoption grew from just 39–46 nodes in 2014 [36] through a slow ramp-up of 2–3% during 2017–2019 before accelerating sharply to 23% by 2021 and 64% by 2025 (Table IV).

The apparent "ASN concentration," with HHI rising from 166 to 4163, is mostly an artefact of Tor adoption rather than cloud provider risk. Hetzner's share *decreased* from 10% to 3.6%.

### D. Tor Infrastructure Resilience

To address the partial observability challenge, we apply both approaches described in Section III-D.

**Distributional bounds.** Table V shows $p_c$ under five Tor distribution scenarios for selected years. Before 2021, when Tor adoption was at most 3%, all scenarios yield near-identical $p_c$, as Tor nodes are too few to affect results. From 2022 onward, as Tor exceeds 50% of the network, scenarios diverge substantially: the 2025 bounds span $p_c = 0.12$ (worst case, all Tor in Tonga) to 0.78 (clearnet only). Excluding the extreme worst case, the plausible range is 0.68 (uniform) to 0.78 (clearnet only), a range of 0.10.

TABLE V: Tor sensitivity bounds: $p_c$ under five relay-distribution assumptions.

| Year | Tor % | Clear. | Prop. | Unif. | Clust. | Worst |
|---|---|---|---|---|---|---|
| 2017 | 3% | 0.92 | 0.92 | 0.90 | 0.92 | 0.92 |
| 2021 | 23% | 0.72 | 0.52 | 0.66 | 0.60 | 0.76 |
| **2022** | **52%** | **0.88** | **0.86** | **0.80** | **0.86** | **0.12** |
| 2023 | 57% | 0.76 | 0.70 | 0.66 | 0.74 | 0.10 |
| **2024** | **64%** | **0.80** | **0.72** | **0.66** | **0.76** | **0.12** |
| **2025** | **64%** | **0.78** | **0.70** | **0.68** | **0.70** | **0.12** |

At country-level granularity, the *worst case* scenario produces dramatically lower $p_c$ ($\approx$ 0.10–0.12) than all other scenarios when Tor exceeds 50%. Concentrating all Tor nodes in a country with a single submarine cable (Tonga) makes the network extremely fragile: cutting one cable disconnects the majority of the network. This extreme bound is informative precisely because it is implausible: in practice, Tor nodes are distributed across dozens of countries, and the more realistic proportional and clustered scenarios show only modest $p_c$ reduction (0.70–0.86 for 2022–2025).

**4-layer mechanistic model.** Table VI compares the 3-layer clearnet-only model with the 4-layer model that includes Tor relay infrastructure. The 4-layer $p_c$ is consistently *higher* than the 3-layer estimate, with $\Delta p_c$ ranging from +0.02 to +0.10 and reaching its maximum when Tor adoption peaks at 64%.

TABLE VI: 3-layer (clearnet) vs. 4-layer (with Tor relay infrastructure) critical failure threshold.

| Year | 3-layer $p_c$ | 4-layer $p_c$ | Tor % | $\Delta$ |
|---|---|---|---|---|
| 2017 | 0.92 | 0.92 | 3% | 0.00 |
| 2021 | 0.72 | 0.76 | 23% | +0.04 |
| **2022** | **0.88** | **0.90** | **52%** | **+0.02** |
| 2023 | 0.76 | 0.84 | 57% | +0.08 |
| 2024 | 0.80 | 0.90 | 64% | +0.10 |
| **2025** | **0.78** | **0.86** | **64%** | **+0.08** |

The mechanism is straightforward: the majority of Tor relay consensus weight is hosted in Germany, France, and the Netherlands, which have extensive submarine cable and land border connectivity. Cable failures severe enough to disconnect peripheral countries do not significantly degrade relay capacity in these well-connected nations. An adversary must remove substantially more cables to simultaneously disrupt both clearnet ASN routing *and* Tor relay infrastructure. The 4-layer $p_c$ falls above the upper end of the distributional bounds (Table V), consistent with relay infrastructure being concentrated in well-connected countries rather than uniformly or adversarially distributed. Tor's primary contribution is *redundancy*, providing an alternative routing layer that meaningfully increases the cable removal threshold. A consensus weight (CW) threshold sensitivity analysis confirms this: sweeping the Tor circuit failure threshold across 30%, 50%, and 70% yields *identical* $p_c$ values for all years, indicating that the result is dominated by physical and ASN layer topology rather than the specific CW parameterization.

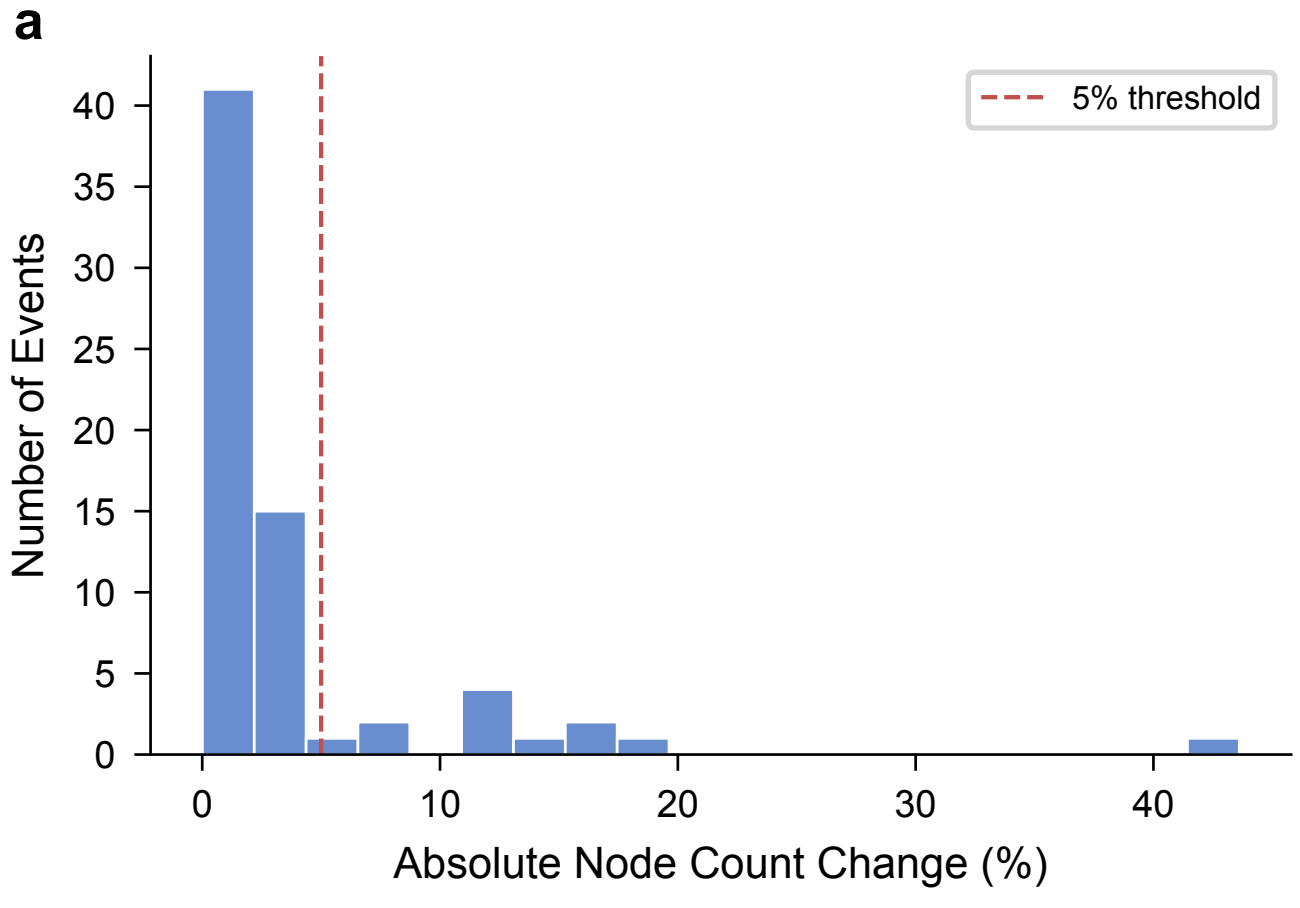

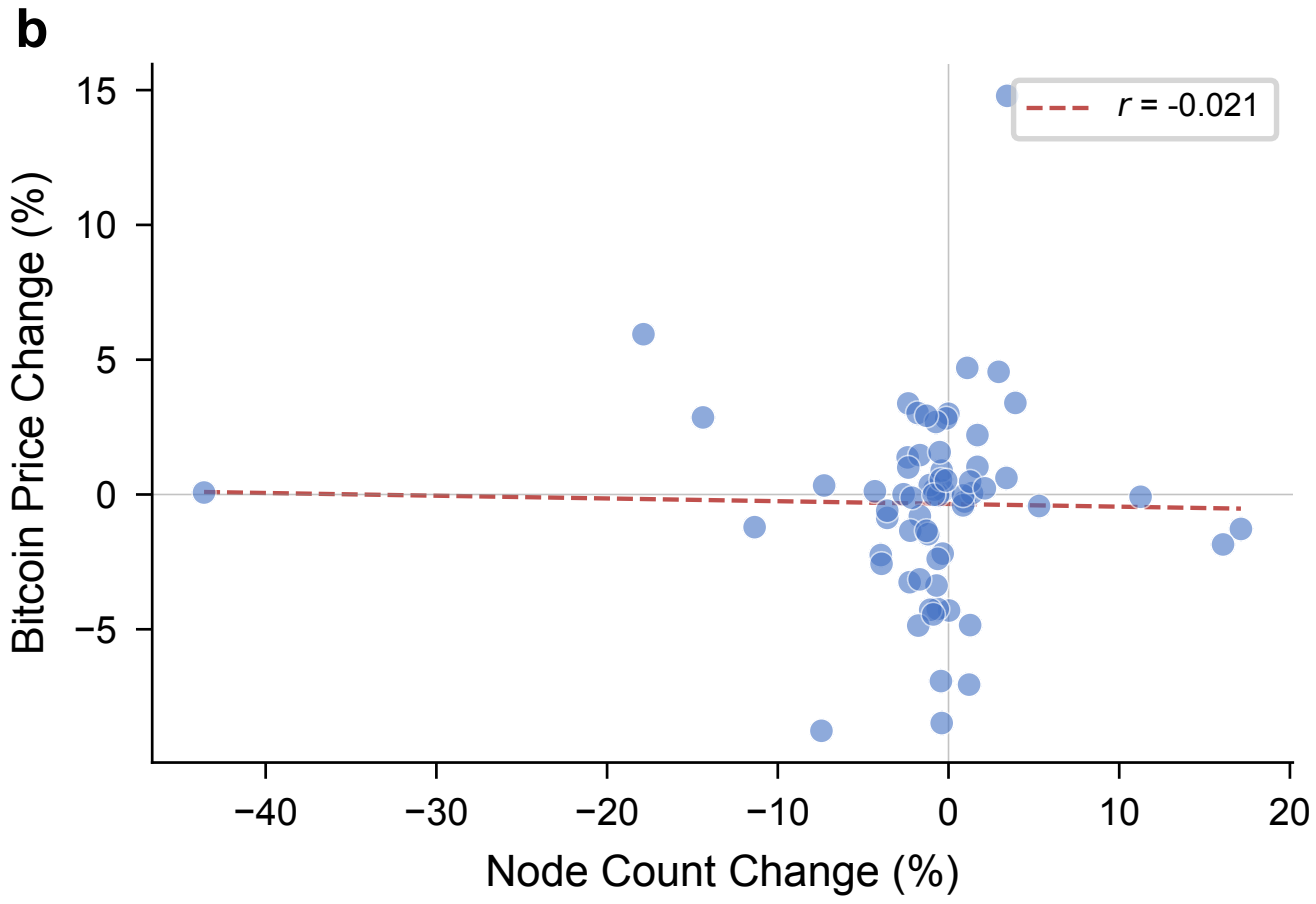

Fig. 4: Empirical validation. (a) 87% of cable events caused <5% node change. (b) Near-zero node-price correlation, $r = -0.02$.

### E. Empirical Validation: Cable Fault Impact

We analyzed all verified cable fault events with matched bitnodes snapshots. Results show 87% of events caused less than 5% node change, with mean impact of −1.5%, median −0.4%, and near-zero price correlation (Fig. 4).

The largest event occurred on March 14, 2024, when seabed disturbances off Côte d'Ivoire simultaneously damaged 7–8 submarine cables, affecting 13 West and Central African countries with an IODA severity score of 11,073. The event caused a $-43.6\%$ regional node drop due to temporary crawler reachability loss that recovered within days, but the affected region hosted only 5–7 Bitcoin nodes, $\approx 0.03\%$ of the global network. The global impact was negligible at $-2.5\%$, within normal fluctuation, consistent with the percolation model: $\approx$ 1–2% of cables removed, far below $p_c \approx 0.80$ for 2024.

Cable events show no detectable effect on Bitcoin's price: the node-change-to-price correlation is $r = -0.02$ ($p = 0.87$, 95% CI $[-0.24, +0.20]$). Infrastructure signals are dwarfed by Bitcoin's daily volatility, consistent with the weak physical-layer coupling our model predicts.

### F. Threshold Sensitivity

Our 10% disconnection threshold is operationally motivated: at this level, block propagation latency increases measurably and mining pools in disconnected countries risk producing orphan blocks. To verify robustness, we compute $p_c$ at 5%, 10%, 15%, and 20% thresholds. Across all years, varying the threshold shifts $p_c$ by $\pm$0.06–0.12: higher thresholds yield lower $p_c$ as less cable removal triggers the threshold, while lower thresholds require more removal. All qualitative findings—the temporal resilience evolution, targeted-vs-random asymmetry, and Tor's amplifying effect—are preserved across all threshold choices.

## V. Discussion

### A. Structural Resilience vs. Buldyrev Cascades

Our results contrast with the catastrophic cascades predicted by Buldyrev et al. [5]. The difference arises from three factors: (1) *partial coupling*, where only 13% of cable events cause $>$ 5% node impact; (2) *geographic diversity*, as nodes concentrate in well-connected countries with extensive submarine and terrestrial connectivity; (3) *partial observability*, since Tor hides the majority of nodes' physical locations.

These factors produce *graceful degradation* rather than catastrophic failure. The order-of-magnitude gap between random and targeted thresholds (Table II, Table III) reflects this structure. Notably, our percolation curves show a *continuous* (second-order-like) transition rather than the abrupt first-order collapse that Buldyrev et al. predict for tightly coupled interdependent networks. The curves exhibit an S-shape: a plateau ($p < 0.20$) with near-zero measurable impact, gradual build-up ($p \approx 0.20$–$0.78$), and an accelerating phase that never jumps discontinuously. Near $p_c$, standard deviation is comparable to the mean (e.g., 2025: mean $= 10.7\%$, std $= 8.4\%$ at $p = 0.78$), and even at $p = 0.98$ failure reaches only $\sim 46\%$, characteristic of continuous transitions rather than catastrophic collapse. This is consistent with theoretical predictions for weakly coupled and structurally correlated multiplex networks: bond percolation on multiplex networks with layer correlation produces continuous rather than discontinuous transitions [37], inter-similarity of degree sequences between layers has the same effect [38], and real-world interdependent networks generally exhibit continuous transitions rather than the catastrophic collapse predicted for idealized random graphs [39]. Bitcoin's multi-homed ASN topology and geographic concentration in well-connected countries attenuate cascade propagation, consistent with the weak-coupling regime [40], [41].

Our analysis focuses on physical and routing layer resilience, but Bitcoin also benefits from protocol-level mechanisms: block relay networks (FIBRE, Falcon), compact block relay (BIP 152) [42], and Blockstream Satellite [43] bypass general-purpose routing or terrestrial infrastructure entirely. These mechanisms add resilience layers our model does not capture, making our $p_c$ estimates conservative.

### B. Tor and Physical Infrastructure

Prior work characterized Tor as providing *censorship resistance* but not *physical resilience* [33], [44], while Malis et al. [45] analyzed submarine cable chokepoints in the context of anonymous communication routing. Our 4-layer model reveals a more nuanced picture. Tor relay infrastructure is heavily concentrated: Germany, France, and the Netherlands together host a majority of consensus weight, reflecting the dominance of high-bandwidth European hosting providers. Critically, these countries have extensive submarine cable and land border connectivity, meaning cable failures predominantly affect peripheral nations with minimal relay capacity.

This geographic correlation between relay infrastructure and cable connectivity explains the gap between 3-layer and 4-layer estimates (Table VI), consistent with findings that relay geographic placement fundamentally shapes Tor's performance and security properties [46], [47]. An adversary targeting submarine cables faces a compound challenge: disrupting connectivity to well-connected European countries is difficult (many redundant paths via both submarine and terrestrial links), and disrupting peripheral cables has limited impact on Tor relay availability. Tor thus provides not only *observability reduction*—an adversary cannot identify which infrastructure to target—but also *structural resilience* through the geographic distribution of its relay network.

This finding qualifies the concern that Tor's unobservability masks hidden fragility. The distributional bounds (Table V) show that under simplified geographic assumptions (proportional, uniform, clustered), Tor adoption modestly *reduces* clearnet $p_c$ by 0.02–0.14, because spreading nodes to additional countries introduces new geographic dependencies. However, the mechanistic 4-layer model (Table VI) consistently shows *higher* $p_c$ than the clearnet-only baseline, because it accounts for the actual Tor relay infrastructure: Tor nodes depend on relay circuits, and these relays are concentrated in well-connected countries that are robust to cable failures. The 4-layer model captures a dependency the distributional scenarios miss, and represents the more realistic estimate.

### C. Limitations

Our distributional bounds and 4-layer model address partial observability—the majority of nodes use Tor with unknown physical locations (Table IV)—but do not fully resolve it. The bounds analysis assumes static Tor node distributions, whereas real operators may relocate in response to events. The 4-layer model uses a current Tor relay snapshot rather than historical data; however, relay geographic concentration has been remarkably stable: Germany alone contributed nearly half of relay bandwidth in 2007 [48] and 35% in 2022 [47], with Germany, France, and the Netherlands consistently ranking among the top four relay countries throughout this period, so the retroactive bias from using current geography is limited. The model assumes a binary consensus weight threshold for circuit failure; a sensitivity analysis across 30%, 50%, and 70% thresholds shows identical $p_c$ values for all years, confirming that results are insensitive to this parameter. Bitnodes crawls reachable listening nodes, representing approximately 40% of the estimated full network; unreachable nodes depend on the same physical infrastructure, so their resilience profile does not differ qualitatively. Our country-level aggregation, while capturing intercontinental chokepoints, may miss finer-grained routing structure; Mühlbauer et al. [49] argue that AS routing policies operate at sub-country granularity. The binary cascade assumption—that nodes in disconnected countries fail immediately—is a simplification, as real failures are probabilistic and some nodes may have alternative connectivity. Finally, cable events and bitnodes snapshots are matched within 7-day windows, so short-lived partitions lasting less than one hour may not be captured.

### D. Implications for Distributed Systems

The multiplex percolation framework applies to any overlay network built on shared physical infrastructure. Resilient overlays should incorporate weak coupling to underlay failures, as Bitcoin's architecture demonstrates. The Tor adoption pattern shows that decentralized systems can develop adaptive resilience without central coordination. Concentration metrics such as HHI and top-5 provider share can be misleading when overlays bifurcate across observable and unobservable layers; resilience analysis should decompose by network type to distinguish genuine risk from observability artefacts.

## VI. Conclusion

We presented the first longitudinal analysis of Bitcoin's physical infrastructure resilience, spanning 11 years and 68 verified cable fault events. Bitcoin is highly resilient to random cable failures but far more vulnerable to targeted attacks (Table II, Table III), and empirical validation confirms this weak physical-layer coupling.

The central finding concerns Tor adoption. Given the present concentration of relay infrastructure in well-connected European countries, our 4-layer model shows that Tor creates a compound barrier to disruption: the full-network $p_c$ consistently exceeds the clearnet-only estimate ($\Delta p_c = +0.02$ to $+0.10$). The 4-layer model captures a dependency that distributional assumptions miss: Tor nodes rely on relay circuits hosted in countries with robust cable connectivity, providing structural redundancy that outweighs added geographic exposure.

Bitcoin's shift to Tor represents self-organized response to regulatory pressure that simultaneously enhances infrastructure resilience, where censorship resistance and physical robustness are complementary rather than competing properties. Country-level mining data from CBECI [35] (2019–2022) and Hashrate Index [34] (2025, 195 countries) suggests that mining geographic diversification has not materially altered infrastructure resilience, consistent with physical cable topology rather than hashrate distribution governing $p_c$. Our multiplex percolation framework provides a foundation for continuous resilience monitoring and extends naturally to other overlay networks built on shared physical infrastructure.

## Acknowledgments

We thank Yong-Yeol Ahn, Alex Arenas, Ulrich Bindseil, Rafael de Oliveira Costa, Claudia Diaz, Gaogao Dong, Bnaya Gross, Matthias Grundmann, Hanna Halaburda, Shlomo Havlin, Rob Jansen, Jameson Lopp, Olaf Maennel, Filippo Radicchi, Alagappan Ramanathan, Louis Shekhtman, Nicole Starosielski, Paul Syverson, and Florian Tschorsch for helpful comments on an earlier draft.